\documentclass[vecphys]{svmult}

\usepackage{graphicx}        
\usepackage[bottom]{footmisc}

\begin{document}

\title*{3D Photoionisation Modelling of NGC~6302}
\author{N.J.~Wright\inst{1} \and
M.J.~Barlow\inst{1} \and
B.~Ercolano\inst{2} \and
T.~Rauch\inst{3}}
\institute{Department of Physics and Astronomy, University College London, Gower Street, London, WC1E~6BT.
\texttt{email:~nwright@star.ucl.ac.uk}
\and Harvard-Smithsonian Center for Astrophysics, 60 Garden Street, Cambridge, MA~02138, USA \and Institut f\"ur Astronomie und Astrophysik, Universit\"at T\"ubingen, Sand 1, 72076 T\"ubingen, Germany}
%
%
\maketitle

\begin{abstract}

We present a three-dimensional photoionisation and dust radiative transfer model of NGC~6302, an extreme, high-excitation planetary nebula. We use the 3D photoionisation code {\sc mocassin} to model the emission from the gas and dust. We have produced a good fit to the optical emission-line spectrum, from which we derived a density distribution for the nebula. A fit to the infrared coronal lines places strong constraints on the properties of the unseen ionising source. We find the best fit comes from using a $220,000$~K hydrogen-deficient central star model atmosphere, indicating that the central star of this PN may have undergone a late thermal pulse.

We have also fitted the overall shape of the ISO spectrum of NGC~6302 using a dust model with a shallow power-law size distribution and grains up to $1.0\mu$m in size. To obtain a good fit to the infrared SED the dust must be sufficiently recessed within the circumstellar disk to prevent large amounts of hot dust at short wavelengths, a region where the ISO spectrum is particularly lacking. These and other discoveries are helping to unveil many properties of this extreme object and trace it's evolutionary history.

\keywords{planetary nebulae: individual: NGC~6302}
\end{abstract}

\section{Introduction}
\label{sec:1}

NGC~6302 is a massive planetary nebula (PN) with a dense circumstellar disk. Estimations of the mass of the ionised nebula ($\sim 0.5-3$~M$_{\odot}$ \cite{mats05}) and the high nebular abundance of nitrogen \cite{alle81} are consistent with it's type I status and imply a massive progenitor \cite{casa00}. The nebula shows a strongly bipolar morphology with a massive equatorial disk, the density of which is sufficient to completely obscure the unobserved central star \cite{mats05}.

Infrared spectroscopy of the central region of NGC~6302 has shown many emission lines from high-ionisation species, in particular that of [Si~{\sc ix}]~3.934~$\mu$m \cite{casa00}, the highest ionisation species found in a PN, as well as lines of [Mg~{\sc viii}] and [Al~{\sc vi}]. These observations point towards a very hot central star with estimates of the surface temperature of up to 430,000~K \cite{ashl88}.

NGC~6302 has a large dust component calculated to have a total mass of 0.05~-~0.05~M$_{\odot}$ \cite{mats05,kemp02}. The ISO spectrum of NGC~6302 \cite{barl98} also shows a wealth of emission features from crystalline dust species, including crystalline silicates \cite{wate96}, crystalline water ice \cite{barl98} and the first extra-solar identification of carbonate features \cite{kemp02}. Despite the large number of oxygen-rich solid state species detected, bands attributed to polycyclic aromatic hydrocarbons (PAH) have been detected \cite{roch86}, usually indicative of carbon-rich material. This dichotomy between O-rich and C-rich material makes NGC~6302 an important case in stellar evolution. Only by compiling a fully three-dimensional photoionisation model of NGC~6302 can a full understanding of this enigmatic object be possible.

\section{The Model}
\label{sec:2}

{\sc mocassin} is a 3D photoionisation code that uses Monte Carlo techniques to solve the thermal equilibrium and ionisation balance equations across an entire grid of arbitrary geometry and density distribution \cite{erco03}. The code allows multiple non-central ionising sources, multiple gas chemistries and has a full dust model incorporating all the gas-dust coupled processes \cite{erco05}.

\subsection{The Nebula Density Structure}

The 3-dimensional nebula model was constructed by fitting the optical emission-line spectrum, particularly the density- and temperature-sensitive line ratios. The chosen optical emission-line spectrum was that of Tsamis et al. (2003) \cite{tsam03}, a deep optical spectrum made with a fixed slit that {\sc mocassin} can simulate when producing model line fluxes for comparison with observations. The IR spectrum was taken from the ISO LWS analysis of Liu et al. (2001) \cite{liu01} and the infrared grating and echelle spectroscopy of Casassus et al. (2000) \cite{casa00}.

\begin{figure}[t]
\begin{minipage}[t]{58mm}
\includegraphics[width=55mm, angle=-1.5]{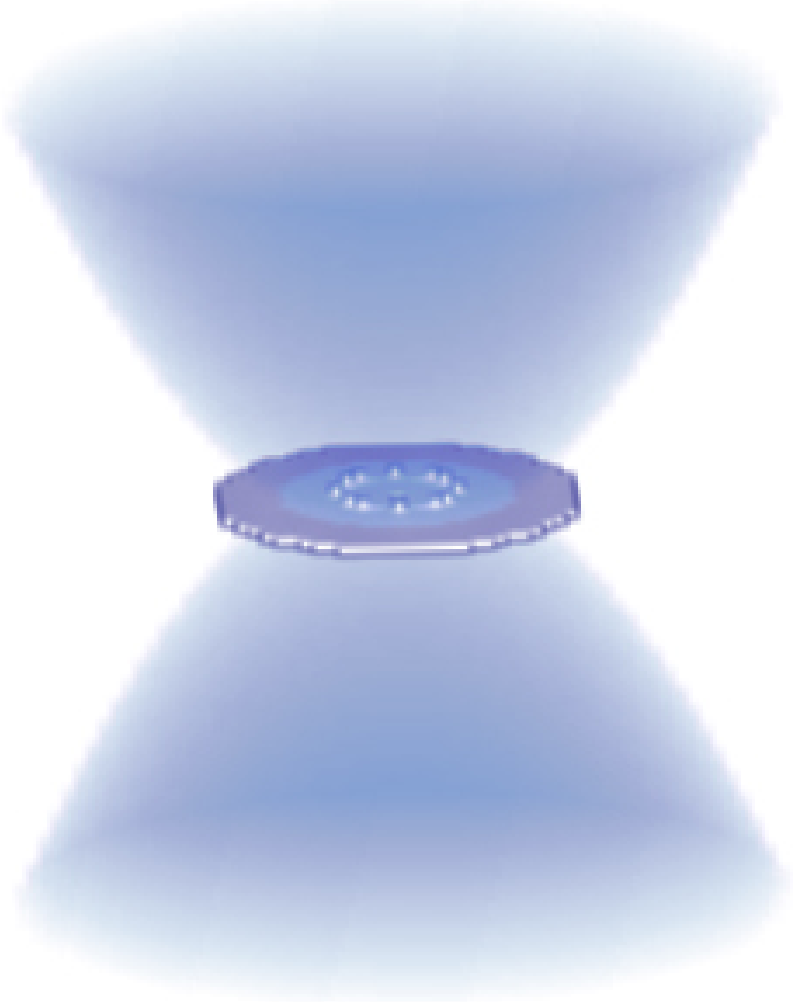}
\caption{Image of the 3D nebula structure used to model NGC~6302.}
\label{imagenebula}
\end{minipage}
\hfill
\begin{minipage}[t]{55mm}
\includegraphics[width=53mm]{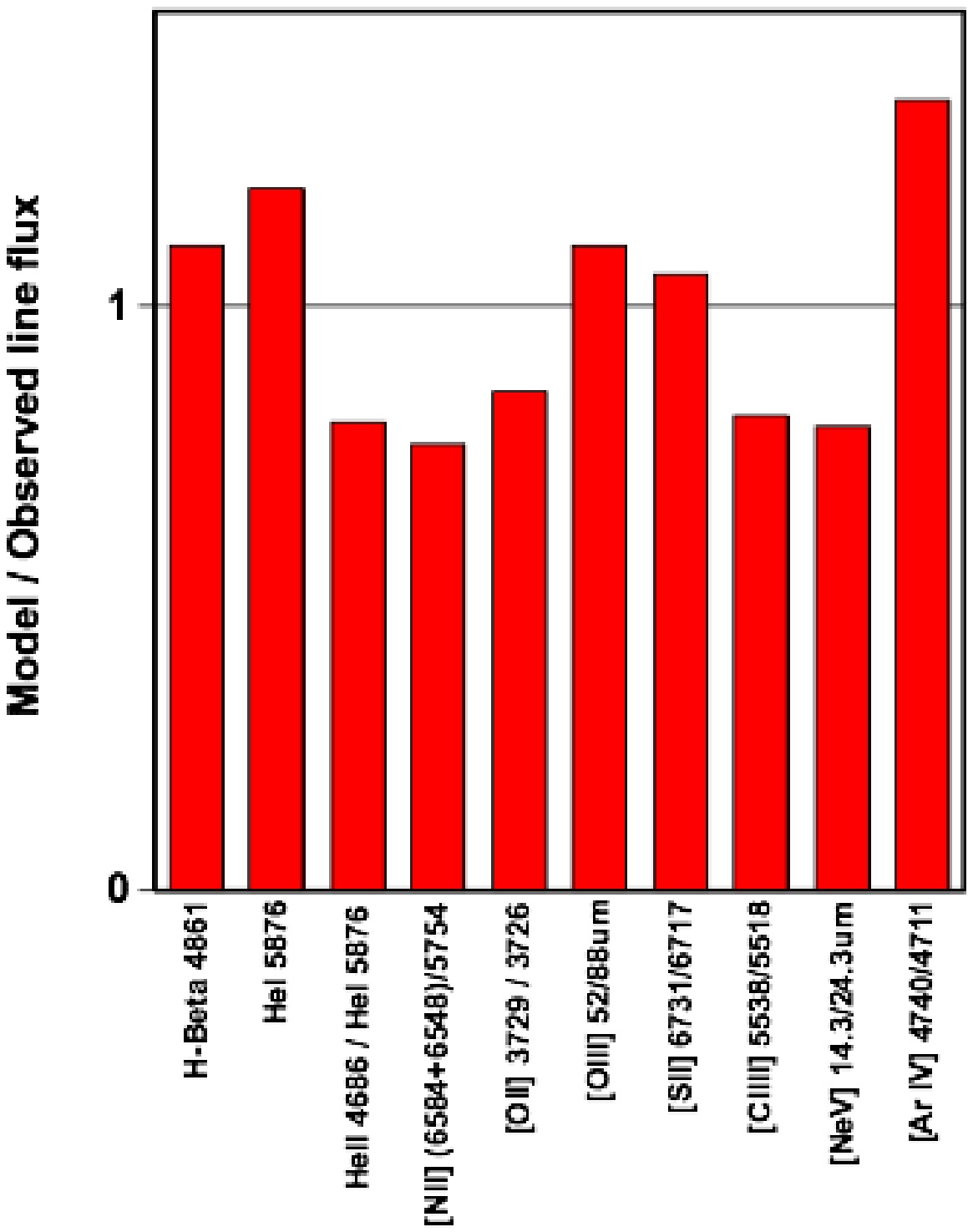}
\caption{Ratios of model-to-observed line fluxes and line ratios for the current best-fit model of NGC~6302.}
\label{lineresults}
\end{minipage}
\end{figure}

We started our models with an hourglass-shaped bipolar nebula with dimensions determined from observations of the nebula and the recently derived distance of 1.04~kpc \cite{mats05, meab05}. This was complemented by a thin circumstellar disk with a radial density dependence \cite{pasc04}. From there we varied the free parameters to achieve a best fit to the observations. Figure~\ref{imagenebula} shows an image of the model nebula showing the diffuse bipolar lobes and the dense circumstellar disk. The nebula lobes are filled and at a constant density of 1000~cm$^{-3}$, while the circumstellar disk follows a radial density dependence, $\rho~\propto~r^{-1}$, which rises to a density of 700,000~cm$^{-3}$ at the inner edge of the disk. The dimensions of the bipolar lobes and the disk are given in Table~\ref{dimensions}.

Figure~\ref{lineresults} shows the current best fit model results for some of the density- and temperature-sensitive line ratios from the optical and IR line spectra. To achieve this fit, a high density contrast between the diffuse bipolar lobes and the dense circumstellar disk was required. The nebula densities derived by Tsamis et al. (2003) for NGC~6302 range from 1380~cm$^{-3}$ from the [O~{\sc iii}] 52 $\mu$m/88~$\mu$m line ratio to 22,450~cm$^{-3}$ from the [Cl~{\sc iii}] $\lambda 5538 / \lambda 5518$ line ratio \cite{tsam03}. Our good fits to both these line ratios show that our extreme densities are a viable method of reproducing the wide range of density-diagnostic line ratios measured in NGC~6302. Binette \& Casassus (1999) \cite{bine99} measure electron densities of $1.4 \times 10^5$~cm$^{-3}$ and $2.3 \times 10^5$~cm$^{-3}$ from the density diagnostic line ratios [Ne~{\sc v}]~24.3~$\mu$m/14.3~$\mu$m and [Ar~{\sc v}]~13.1~$\mu$m/7.91~$\mu$m respectively. The latter of which is a good density diagnostic in the range 10,000-300,000~cm$^{-3}$ and is highlighting the extremely high densities in the circumstellar disk which we have modelled and also achieve a good match for.

The derived density structure that offers us the best fit to the observations leads us to a total nebula mass of 0.428~M$_{\odot}$, with approximately 40~\% of the mass in the circumstellar disk. This disk mass of 0.160~M$_{\odot}$ is far short of the 3~M$_{\odot}$ derived by Matsuura et al. (2005) from dust masses from the infrared SED and using a dust-to-gas ratio of 0.01 . It should be noted that our model masses are only those that are necessary to reproduce the observations, and that, because the disk is optically thick when viewed radially, we are obviously underestimating this mass by potentially an order of magnitude. If the outer radius of the disk is extended to $3.0 \times 10^{17}$~cm (increasing the mass by a factor of 5.0) we find no appreciable difference in the produced spectrum, indicating that large amounts of neutral or molecular material could be hidden within the disk, without affecting the ionisation structure of the nebula.

\begin{table}[t]
\caption{Dimensions used in the 3D density distribution for the model of NGC~6302}
\label{dimensions}
\begin{tabular}{lc p{0.5cm} lc}
Lobe dimensions &&& Disk dimensions \\
\cline{1-2} \cline{4-5}
Parameter & Value && Parameter & Value \\
\cline{1-2} \cline{4-5}
\vspace*{-0.25cm} \\
Length of lobe & 6.8 $\times 10^{17}$ cm                             && Inner radius & 1.0 $\times 10^{16}$ cm   \\
Waist radius of lobe & 1.0 $\times 10^{16}$ cm                    && Outer radius & 1.0 $\times 10^{17}$ cm   \\
Maximum radius of lobe & 3.1 $\times 10^{17}$ cm             && Disk height & 5.0 $\times 10^{15}$ cm    \\
Volume of lobes (both) & 3.2 $\times 10^{53}$ cm$^3$      && Volume of disk & 2.24 $\times 10^{51}$ cm$^3$  \\
Density of lobes & 1000 H-atoms cm$^{-3}$                          && Density profile & $\rho \propto r^{-1}$  \\
Mass of lobes (both) & 0.268 M$_{\odot}$                              && Mass of disk & 0.160 M$_{\odot}$   \\
\cline{1-2} \cline{4-5}
\end{tabular}
\end{table}

Initially we have not attempted to model any density inhomogeneities or to reproduce the many small-scale structures seen in the lobes of NGC~6302 \cite{mats05}. This is mainly due to a lack of spatial information on the gas properties but also because of the importance of compiling as simple a model as possible. We have also maintained a completely homogeneous gas chemistry throughout the model nebula.  The chemical abundances used in the model are taken from the analysis of Tsamis et al. (2003) \cite{tsam03} and Casassus et al. (2000) \cite{casa00} and incorporates 14 elements.

\subsection{The IR Coronal Lines}

While the central ionising source of NGC~6302 has never been observed, the emission from the IR coronal lines puts a constraint on the ionising flux. Of all the high-ionisation lines observed, those of Mg$^{+7}$ and Si$^{+8}$ \cite{casa00} are of the highest stages and, if predominantly photo-ionised, will put a strong constraint on the high-energy end of the ionising spectrum.

The first observations of IR coronal lines in NGC~6302 were the prominent [Si~{\sc vi}] and [Si~{\sc vii}] lines \cite{ashl88}. Previous attempts with one-dimensional photoionisation codes have been unable to accurately reproduce these lines \cite{lame91} leading to suggestions that these lines may arise from collisionally-excited material \cite{rowl94}. Many of the arguments for shock-excited material originated from the observation of $\sim$500~kms$^{-1}$ wings on the [Ne~{\sc v}]~$\lambda$3426 line \cite{meab80}. However, similar wings were not seen by Casassus et al. (2000) in their observations of infrared coronal lines, despite similar excitation potentials. The origin of the [Ne~{\sc v}]~$\lambda$3426 wings is now thought to be instrumental \cite{meab05}. We are therefore attempting to fully reproduce all the high-excitation lines by means of photoionisation alone.

To model emission from the central star we use non-LTE stellar atmospheres calculated using the T\"ubingen NLTE Model Atmospheres Package \cite{rauc03}. We have used both the solar abundance models that incorporate all elements up to nickel, and the hydrogen-deficient models which use "typical" PG~1159 abundance ratios of He:C:N:O~=~33:50:2:15 by mass. For both of these we have varied both the effective temperature and the gravity of the central star in attempting to fit the coronal lines. We also varied certain nebula parameters such as the disk density and height, that might effect the ionisation structures of these elements.

Our models reveal that the majority of the nebula parameters have little effect on these highly-ionised ions. The effective temperature of the central star has by far the strongest influence on the ionisation structure, yet by using a solar abundance model we were unable to reproduce the highest ionisation stages without reaching temperatures in excess of 400,000~K, where many of the model atmosphere calculations may not be applicable \cite{rauc03}. Even at lower temperatures, the ionisation structure that a solar abundance atmosphere creates does not allow accurate reproduction of even the lower-ionisation lines. This can be seen in Table~\ref{highlines} for the ionisation structures of magnesium and silicon, where the lower ionisation lines (from Mg$^{+4}$ and Si$^{+5}$) are over-predicted, while no flux is seen in the high-ionisation lines (from Mg$^{+7}$ and Si$^{+8}$).

However, models using H-deficient central star atmospheres produced ionisation structures that matched the observations much better and were able to reproduce the highest ionisation lines at much lower temperatures (see Table~\ref{highlines}). Order of magnitude matches can be obtained for nearly all the lines using a 220,000~K model atmosphere, reproducing the ionisation structures of these elements much better.

\begin{table}[t]
\caption{Observed and modelled line strengths for the high-ionisation IR coronal lines of NGC~6302. All line strengths given relative to H$\beta \times 10^{-3}$. Observations from Casassus et al. (2000).}
\begin{tabular}{l p{0.5cm} c p{0.75cm} c p{0.75cm} c}
\hline
$$ &&&& Solar abundances && H-deficient \\
ID $\lambda$ && Observations && 220 kK && 220 kK \\
\hline
$$[Mg {\sc v}] 5.60$\mu$m && 2.19 && 59.2 & & 5.23 \\
$$[Mg {\sc vii}] 5.51$\mu$m && 1.68 && 1.06 & & 0.60 \\
$$[Mg {\sc viii}] 3.03$\mu$m && 0.234 && 0.0 & & 0.58 \\
$$[Si {\sc vi}] 1.96$\mu$m && 3.04 && 34.0 & & 2.23 \\
$$[Si {\sc vii}] 2.48$\mu$m && 2.66 && 0.09 & & 0.22 \\
$$[Si {\sc ix}] 3.93$\mu$m && 0.00389 && 0.0 & & 0.0318 \\
\hline
\end{tabular}
\label{highlines}
\end{table}

\begin{figure}[t]
\begin{center}
\includegraphics[width=70mm, angle=270]{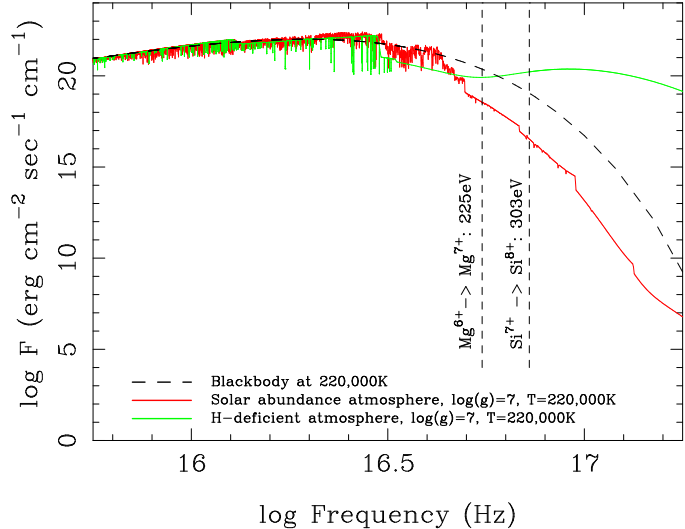}
\caption{Spectral energy distributions of two of the solar abundance and hydrogen-deficient stellar atmospheres used in our models. Both stellar atmospheres have T$_{\star}=220,000$~K and log~$g = 7$. The ionisation potentials of the Mg$^{+6}$ and Si$^{+7}$ ions are indicated. A 220,000~K blackbody spectrum is shown for comparison.}
\label{atmos}
\end{center}
\end{figure}

The difference between a solar abundance stellar atmosphere and a hydrogen-deficient stellar atmosphere is most apparent at high energies where the lack of opacity due to hydrogen causes a greater flux of high-energy photons. This can be seen in Figure~\ref{atmos} where we show the spectral energy distributions of a solar-abundance and a hydrogen-deficient stellar atmosphere. The ionisation potentials of Mg$^{+6}$ (225eV) and Si$^{+7}$ (303eV), are also indicated, and it can be seen that a hydrogen-deficient stellar atmosphere has a significantly higher flux at these energies than a solar-abundance stellar atmosphere. It is this difference in the spectral energy distribution of the ionising spectrum that allows a much better fit to the ionisation structure apparent in NGC~6302.

\section{The Dust Model}

{\sc mocassin} has a complete dust model, with the dust competing with the gas for UV photons and interacting directly with it via gas-grain collisions \cite{erco05}. It allows any mix of dust species and grain sizes in multiple density distributions. NGC~6302 is bright in the infrared and the dust distribution has beeen shown to be complex \cite{mats05}, with the overall broad shape of the SED indicating a large range of dust temperatures. These two factors make it important to model the dust using a full 3-dimensional dust radiative transfer code such as {\sc mocassin}.

\begin{figure}[t]
\begin{center}
\includegraphics[width=70mm]{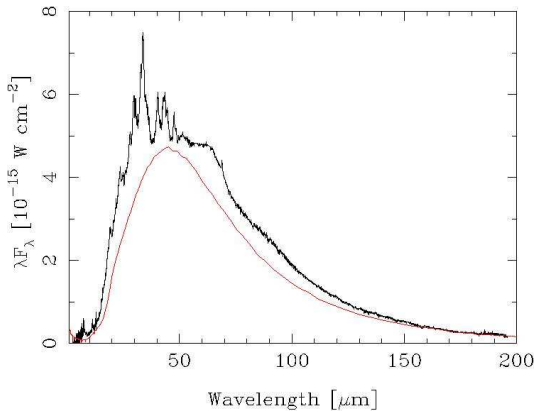}
\caption{{\sc mocassin} fit to the ISO SED of NGC~6302.}
\label{sedfit}
\end{center}
\end{figure}

We have initially used a model with a single dust species using the astronomical silicates model of Draine (2003), which includes optical constants covering the full wavelength range of X-rays to the sub-mm \cite{drai03}. In a full treatment of the radiative transfer of dust it is important to have a continuous coverage of optical constants to accurately treat the absorption by the grains of UV photons and the re-emission in the IR. We started with a simple MRN grain-size distribution using a$_{min} = 0.04\mu$m and a$_{max} = 0.4\mu$m and a slope of $-3.5$. We found however that a better fit could be obtained with larger grains and a shallow slope, producing more larger grains and less small grains. Figure~\ref{sedfit} shows our current single-species fit to the ISO SED.

This fit to the SED uses a power-law grain size distribution with A$_{min} = 0.06\mu$m and A$_{max} = 1.0\mu$m with a slope of $-2.5$. The presence of large grains up to $10 \mu$m in size has been implied from the sub-mm continuum \cite{hoar92} and the presence of such a dense circumstellar disk provides a stable environment where such large grains could be built up.

A better fit to the shape of the SED could be gained by modelling two separate populations of dust grains, one consisting of large grains in the circumstellar disk and the other with a more standard grain-size distribution that might exist in the bipolar lobes of the nebula. The inclusion of multiple grain distributions, with different grain sizes and different chemistries in separate 3-dimensional density distributions is a unique and versatile feature of {\sc mocassin} and will allow the observations of NGC~6302 to be much more accurately reproduced and studied.

The ISO spectrum of NGC~6302 \cite{barl98} also shows a wealth of emission features from crystalline dust species, including crystalline silicates \cite{wate96,mols01} and crystalline water ice \cite{barl98}. Two previously unidentified features at 62 and 90~$\mu$m were also identified as belonging to the carbonates calcite and dolomite \cite{kemp02}, though it is thought that carbonates might not be able to form in these environments \cite{ferr05}.

We are in the process of compiling a full database of dust species suitable for use in photoionisation and radiative transfer codes such as {\sc mocassin}. We intend to gather optical constants for all dust species that have been observed in NGC~6302 or could possibly exist in such environments and attempt to reproduce the detailed emission features seen in the ISO spectrum. A fit to the spectral energy distribution will allow us to derive dust masses for all the species present and also determine if the oxygen-rich and carbon-rich species are spatially separate, as some evolutionary scenarios have suggested.

\section{Conclusions}

We have presented initial results from our three-dimensional photoionisation and dust radiative transfer model of NGC~6302 using the 3D photoionisation code {\sc mocassin}. Our fit to the emission-line spectra has resulted in a high density contrast between a large pair of diffuse bipolar lobes and a thin circumstellar disk where densities reach $700,000$~cm$^{-3}$. This density distribution supports the multiple density diagnostics of NGC~6302 that cover a large range of densities.

Our fit to the infrared coronal lines comes using a $220,000$~K hydrogen-deficient stellar atmosphere model. This is one of the lowest temperatures derived for the unseen central star of this object, which is only possible using a hydrogen-deficient central star model, implying that the central star of NGC~6302 may have undergone some sort of late thermal pulse.

Our dust model includes a fit to the overall shape of the SED using a dust grain size distribution that extends up to $1.0\mu$m. The future inclusion of multiple crystalline silicate species in separate dust density distributions will allow us to reproduce the ISO spectrum and determine chemical properties of the dust grains around this object. These and other discoveries are helping to unveil many interesting properties of this interesting object.

\end{document}